\documentclass[letterpaper, 10 pt, conference]{ieeeconf}  
\IEEEoverridecommandlockouts 
\usepackage{multirow}
\usepackage{bbold}
\usepackage{ifpdf}
\usepackage[normalem]{ulem}
\usepackage{caption}
\usepackage{subcaption}%
\usepackage[cmex10]{amsmath}
\usepackage{amssymb}
\usepackage{verbatim}
\usepackage{algorithm}
\usepackage{algorithmic}
\usepackage{array}
\usepackage{latexsym}
\usepackage{color}
\usepackage{textgreek}
\usepackage{subscript}
\usepackage{rotating}
\usepackage{booktabs,multirow}
\usepackage{mdframed}	
\usepackage{tabularx}
\usepackage{amsmath}
\usepackage{makecell}
\usepackage{tabto}
\usepackage{mathrsfs}
\usepackage{url}
\usepackage{cite}
\usepackage{listings}
\usepackage{array}
\usepackage[table]{xcolor}
\usepackage{ragged2e}
\usepackage{booktabs}

\usepackage{graphicx}
\usepackage[cmex10]{amsmath}
\usepackage{amssymb}
\usepackage{bbold}
\usepackage{float}
\usepackage[cmex10]{amsmath}
\usepackage{amssymb}
\usepackage{verbatim}
\usepackage{array}
\usepackage{latexsym}
\usepackage{color}
\usepackage{tabularx} 
\usepackage{textgreek}
\usepackage{subscript}
\usepackage{rotating}
\usepackage{booktabs,multirow}
\usepackage{mdframed}	
\usepackage{tabularx}
\usepackage{amsmath}
\usepackage{makecell}
\usepackage{tabto}
\usepackage{mathrsfs}
\usepackage{url}
\usepackage{cite}
\usepackage{listings}
\usepackage{array}
\usepackage[table]{xcolor}

\usepackage{xcolor} 

\definecolor{lansu}{rgb}{0.4, 0.5, 0.8} 

\usepackage{soul,xcolor}
\usepackage{algorithmic}
\usepackage{array}
\usepackage{soul}
\sethlcolor{green}
\usepackage{url}
\usepackage{pifont}
\usepackage{xcolor}
\usepackage{multirow}
\usepackage{graphicx} 
\usepackage{authblk}

\title{\LARGE \bf Advancing Autonomous Vehicle Safety: A Combined Fault Tree Analysis and Bayesian Network Approach}

\author{
    Lansu Dai and 
    Burak Kantarci \thanks{Lansu Dai and Burak Kantarci are with the School of Electrical Engineering and Computer Science, University of Ottawa, Ottawa, ON, Canada. Emails: \{ldai095, burak.kantarci\}@uottawa.ca.}
    
}
\begin{document}

\maketitle
\thispagestyle{empty}
\pagestyle{empty}

\begin{abstract}
    This paper integrates Fault Tree Analysis (FTA) and Bayesian Networks (BN) to assess collision risk and establish Automotive Safety Integrity Level (ASIL) B failure rate targets for critical autonomous vehicle (AV) components. The FTA-BN integration combines the systematic decomposition of failure events provided by FTA with the probabilistic reasoning capabilities of BN, which allow for dynamic updates in failure probabilities, enhancing the adaptability of risk assessment. A fault tree is constructed based on AV subsystem architecture, with collision as the top event, and failure rates are assigned while ensuring the total remains within 100 FIT. Bayesian inference is applied to update posterior probabilities, and the results indicate that perception system failures (46.06 FIT) are the most significant contributor, particularly failures to detect existing objects (PF5) and misclassification (PF6). Mitigation strategies are proposed for sensors, perception, decision-making, and motion control to reduce the collision risk. The FTA-BN integration approach provides dynamic risk quantification, offering system designers refined failure rate targets to improve AV safety.
\end{abstract}

\section{Introduction}
With the increased use of autonomous vehicles (AVs), it has become critical to ensure the safety of AVs in complex and dynamic environments so that they can accurately perceive, predict, and respond to diverse scenarios to mitigate risks \cite{SaftyAVChallenge}. According to ISO 26262, risk is defined as the combination of the probability of harm occurring and the severity of its consequences \cite{ISO26262}. In the context of AVs, risk arises from system limitations, component failure, and external environmental conditions that may lead to undesirable outcomes, such as collisions. Mitigating risks are important to ensure the safe operation of AVs. This highlights the importance of risk assessment frameworks to address the uncertainty and complexity of the real-world environment. 

Fault tree analysis (FTA) is a widely used technique in safety analysis, particularly in complex systems \cite{FTAsurvey}. It provides a structured approach to identify potential hazards and assessing the likelihood of failure. AVs rely on multiple interconnected systems, they cooperate to ensure the safe and reliable operation. FTA provides a systematic approach to analyze failures across these systems. However, traditional FTA has limitations in handling complex interactions and dependencies in systems \cite{FTAlimitation}. It assumes that the system components fail independently and with static probabilities, making dynamic relationships difficult to model. This challenge becomes evident in autonomous vehicles, where subsystem interactions continuously evolve. 

Bayesian networks (BNs) excel at modeling uncertain events and capturing dependencies between components, enhancing the overall accuracy and flexibility of risk assessment \cite{BNbook}. BNs enable probabilistic inference and dynamic updates based on new evidence. Integrating FTA with BN provides a comprehensive risk assessment framework, combining FTA's ability to decompose failure events systematically with BN's ability to model complex dependencies. While FTA-BN integration has been successfully applied to safety-critical domains, such as unmanned aerial vehicles \cite{UAV} and radio altimeter systems \cite{RadioSystem}, its application to AV risk assessment at the subsystem level remains underexplored. Bhavsar et al. \cite{ProbabilityBhavsar} use traditional FTA to analyze AV systems operating in mixed traffic streams. However, existing analyses rely on static failure probabilities modeling the AV system as a whole. Therefore, state of the art remains the open issue of capturing the intricate interdependencies between its subsystems. This gap leaves a critical aspect of AV safety unexplored, that how failures in individual subsystems contribute to overall system risk and how these risks evolve over time. 

This study addresses the limitations of traditional FTA by incorporating Bayesian inference to derive target failure rates for basic events, which provides valuable guidance for system designers in ensuring AV safety compliance. This paper contributes to risk-based AV safety assessment by focusing on subsystem-level analysis and utilizing BN to dynamically update failure probabilities. The main contributions of this paper can be summarized as follows: 
\begin{enumerate}
    \item We define target failure rates for AV subsystems that align with Automotive Safety Integrity Level (ASIL) B under ISO 26262 functional safety standards \cite{ISO26262}, ensuring that AV system designers can use these failure rate thresholds as guidelines for safety compliance.
    \item We decompose the AV system into subsystems based on its architecture, which includes sensors, perception, decision-making, and motion control systems. This structured approach allows for more targeted risk mitigation strategies and system improvements.
    \item We identify the most significant contributors to collision risk and map critical failure pathways, offering insights for prioritizing safety enhancements and resource allocation in AV development.
\end{enumerate}

Our quantitative analysis demonstrates that perception system failures are the primary contributors to AV collision risk, accounting for 46.06 FIT, almost half of the total 100 FIT failure rate. Specifically, failures in detecting existing objects (PF5), misclassification (PF6), and delayed response to obstacles (DMF2) are identified as critical factors. The rest of the paper is organized as follows. Section II reviews the related work and highlights the novelty of this study. Section III details the methodologies, including FTA and BN integration.  The experimental results and discussion are presented in Section IV. Finally, Section IV concludes with future directions.

\section{Related Work}
FTA has been widely used to evaluate AV safety by identifying potential failure points and assessing associated risks. Bhavsar et al. \cite{ProbabilityBhavsar} employ FTA to analyze vehicular and infrastructure components in mixed-traffic environments, identifying critical failure points in AVs. While promising, their study can further be improved by incorporating dynamic failure probabilities and accounting for real-time updates based on dynamic traffic conditions. Chen et al. \cite{FTAinAVChen} use FTA to examine control transitions in Level 2 and 3 AVs, identifying key failure sources in operational design domains and human-machine interactions. Their study highlights the significant role of human factors in AV safety, particularly in takeover scenarios. Expanding their study to include broader system failures and incorporating in-depth quantitative methods for handling dependencies could further strengthen the findings. Li et al. \cite{FTAinAVLi} integrate FTA with Failure Modes and Effects Analysis (FMEA) to build a fault database. This structured method improves failure diagnosis in electric AVs. The work could be further improved by incorporating real-time adaptability and capturing dependencies between events to provide a more comprehensive and dynamic risk assessment framework.

These studies demonstrate the effectiveness of FTA in AV risk assessment but also highlight its limitations, particularly its static nature and inability to model interdependencies dynamically. To address these challenges, integrating FTA with BN has been explored in other domains, such as radio altimeter systems \cite{RadioSystem} and unmanned aerial vehicles \cite{UAV}. Gao et al. \cite{RadioSystem} used an FTA-BN approach to fault diagnosis in radio altimeter systems. They highlight the advantages of using BN to capture dependencies between system components and update failure probabilities in real-world scenarios. Similarly, Xiao et al. \cite{UAV} demonstrate the effectiveness of BN in UAV safety analysis, by modeling statistical dependencies among UAV components and dynamically updating risk assessments based on real-time flight data.

Despite its success in these fields, FTA-BN integration remains underdeveloped in AV safety research. Most existing studies, including Bhavsar et al. \cite{ProbabilityBhavsar} could benefit from incorporating Bayesian inference. This gap provides an opportunity to apply the FTA-BN framework to AV risk assessment, particularly in modeling subsystem interactions and dynamic risk probabilities. This study aims to bridge that gap, using FTA-BN integration to develop a more adaptive framework for improving AV safety.

\section{Methodology}
\subsection{Fault Tree Analysis (FTA)}
The fault tree is a top-down approach for modeling the pathways leading to a specific system failure or undesired event. It has two main types of nodes: events (basic, intermediate, or top events) and gates (AND/OR) to define logical relationships. 

FTA supports both qualitative and quantitative analysis \cite{FTAsurvey}. The qualitative FTA uses minimal cut sets to identify a system's vulnerabilities without requiring numerical data. The minimal cut sets represent the smallest combination of basic events that can cause the top event. Formally, a minimal cut set \(C_{min}\) is a set of basic events {\(E_1, E_2, \dots, E_n\)}: \(\phi(C_{min}) = 1\) and \(\forall C' \subsetneq C_{min}\), \(\phi(C') = 0\), where \(\phi\) is the structure function representing the logical relationships in the fault tree. 

Quantitative FTA calculates the failure probability of the undesired event occurring by propagating the failure probabilities of the basic events through gates. For an AND gate, the output probability is calculated as the product of the probabilities of all input events. For an OR gate, the output probability is computed as the complement of the product of the complements of the input probabilities. The failure probability of the top event is calculated by iteratively applying these formulas from the bottom to the top of the fault tree. Conducting both qualitative and quantitative FTA provides a comprehensive understanding of system risks and failure pathways, enabling informed risk management decisions.

\subsection{Bayesian Network (BN)}
Bayesian Networks have both forward and backward analysis. The forward analysis calculates the probability of occurrence of any node in the network based on the prior probability of the parent node and the conditional dependence of each node. This provides the prediction of outcomes given known input. The backward analysis focuses on the computation of the posterior probability of any given set of variables given evidence. This allows for reasoning and diagnostics based on known outcomes \cite{CompareFTABN}. 

BNs are widely used to model uncertainty, make inferences, and predict outcomes based on partial information. Therefore, they are useful for modeling uncertainty in autonomous vehicles, as dependencies between variables can be probabilistic and dynamic. BNs can incorporate evidence, such as the failure rate of specific nodes, and then use Bayes' theorem to calculate the posterior failure probability. This makes them a valuable tool for risk assessment.

\subsection{Integrating FTA with BN}

\begin{figure}[htbp]
\centerline{\includegraphics[scale=0.55]{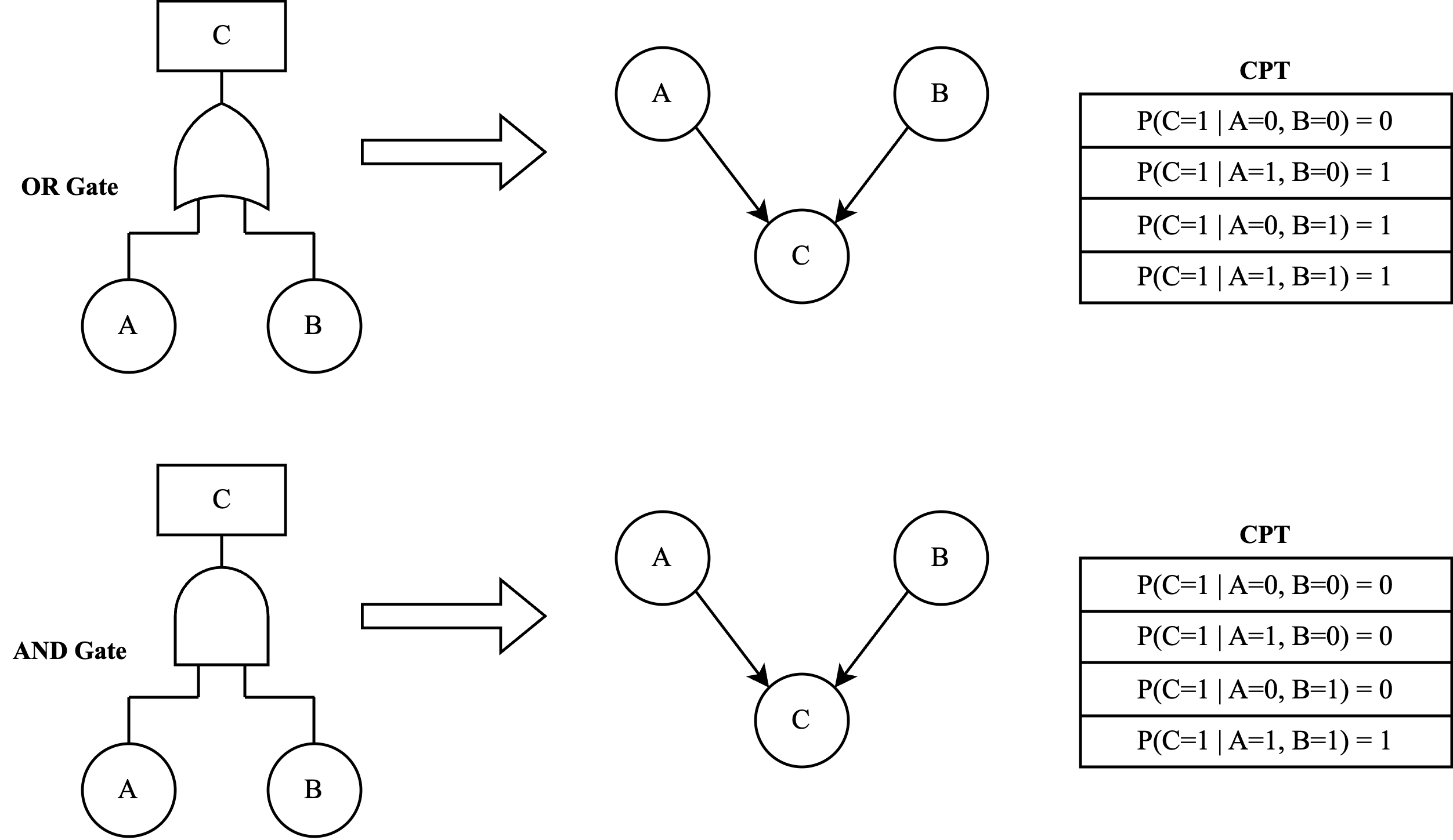}}
\caption{Transformation of AND and OR gates from Fault Tree to Bayesian Network}
\label{fig1}
\end{figure}

Integrating BN with FTA enhances risk assessment by capturing probabilistic dependencies between system components. Any fault tree can be converted into a corresponding BN by creating a binary BN node for each event in the fault tree \cite{MapFTtoBN}. According to Xiao et al. \cite{ExFTABN1}, the conversion rule from fault tree to BN can be considered as two parts: graphical and numerical mapping.

Let \(B = \{ B_1, B_2, \dots, B_n \}\) and \(I = \{ I_1, I_2, \dots, I_m \}\) represent the set of basic events and intermediate events in FTA, respectively. Let \(T\) represent the top event in FTA. The graphical mapping to BN is:

\begin{gather}
B \xrightarrow{\mathcal{M}_G} Pa = \{ Pa_1, Pa_2, \dots, Pa_n \} \\
I \xrightarrow{\mathcal{M}_G} N = \{ N_1, N_2, \dots, N_m \} \\
T \xrightarrow{\mathcal{M}_G} C  
\end{gather}

where \(Pa\) is the set of parent nodes, \(N\) is the intermediate nodes, and \(C\) is the child node in BN. For the numerical mapping, let \(P(B_i)\) represent the occurrence probability of a basic event \(B_i\) and \(G\) represent a logic gate (e.g., AND, OR) in the fault tree.
\begin{gather}
P(B_i) \xrightarrow{\mathcal{M}_N} \text{Prior}(Pa_i) \\
G(B_1, B_2, \dots, B_k) \xrightarrow{\mathcal{M}_N} \text{CPT}(C|N_1, N_2, \dots, N_k)
\end{gather}
where \(\text{Prior}(P_i)\) is the prior probability in BN and \(\text{CPT}(C|N_1, N_2, \dots, N_k)\) is the conditional probability table in the BN corresponding to the logic of gate \(G\).
The translation rule can be summarized as:
\begin{equation}
\text{FTA}( B, I, T, P(B), G ) \xrightarrow{\mathcal{M}} \text{BN}( Pa, N, C, \text{Prior}(Pa), \text{CPT} )
\end{equation}
where \( \mathcal{M} = \mathcal{M}_G \cup \mathcal{M}_N \) represents the combined graphical and numerical mapping functions. Fig.~\ref{fig1} shows the transformation of the two-states AND and OR gates rule. These CPTs form the foundation for probabilistic reasoning in BN derived from fault trees, allowing us to calculate the likelihood of events under uncertainty and update probabilities when new evidence is introduced. Bayes' Theorem is used to incorporate evidence updates.
\begin{equation}
P(E_i \mid \text{Evidence}) = \frac{P(\text{Evidence} \mid E_i) \cdot P(E_i)}{P(\text{Evidence})}
\end{equation}
where \(P(E_i \mid \text{Evidence})\) is the updated probability of event \(E_i\) after considering the evidence, \(P(\text{Evidence} \mid E_i)\) is the probability of observing the evidence, given the event \(E_i\) is true, and \(P(E_i)\) is the occurrence probability before considering the evidence. \(P(\text{Evidence}\) is the marginal probability of the evidence, which can be computed as:

\begin{equation}
P(\text{Evidence}) = \sum_j P(\text{Evidence} \mid E_j) \cdot P(E_j)
\end{equation}
where \(E_j\) represents all possible events in the network.

\subsection{Construction of FT}
\begin{figure}[htbp]
\centerline{\includegraphics[scale=0.2]{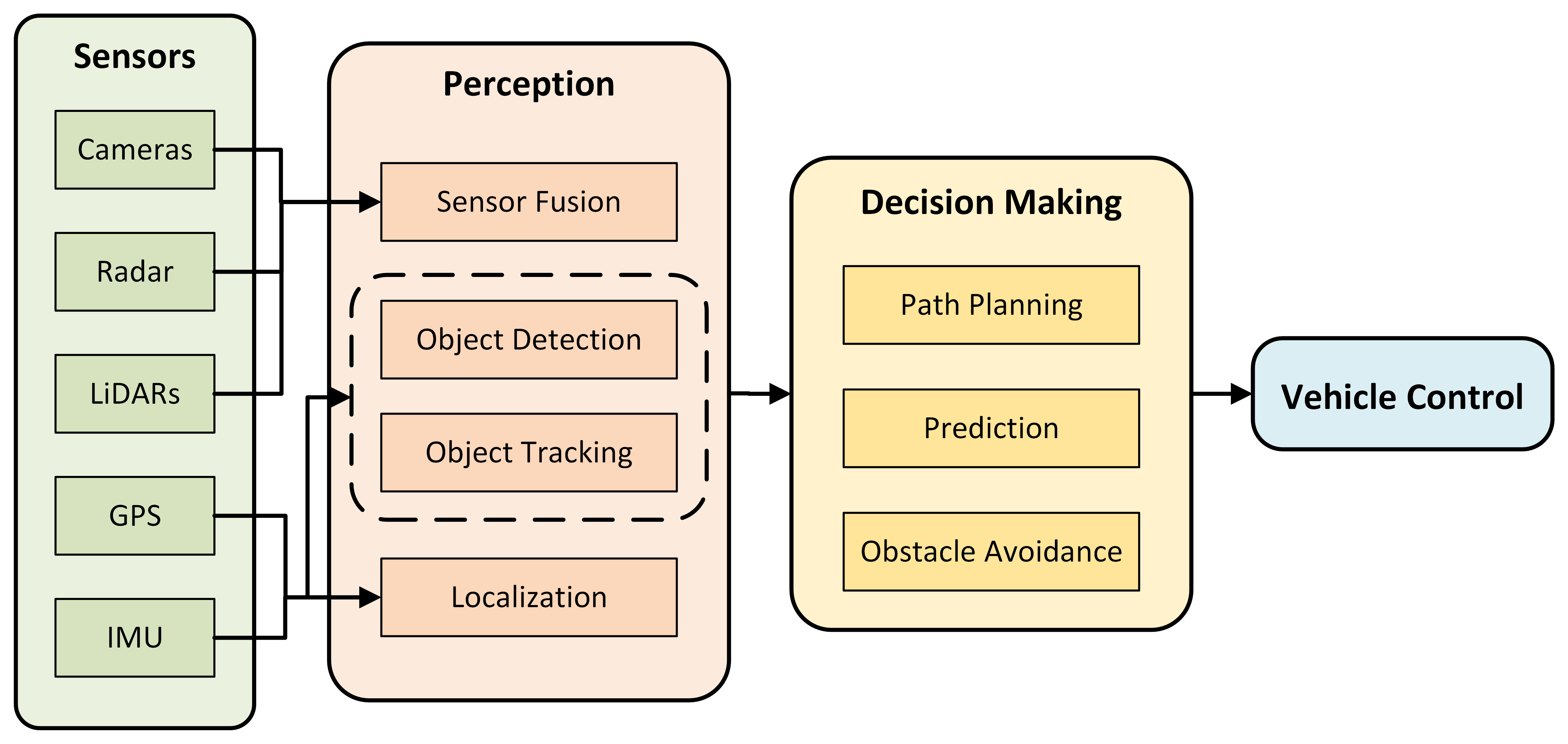}}
\caption{Subsystem Architecture of AV Systems} 
\label{fig2}
\end{figure}

The fault tree for AV collision risk is structured based on the core subsystems of an autonomous vehicle, which includes sensors, perception system, decision-making system, motion control system, and external interactions \cite{Architecture, AVsystem}. The subsystem architecture of AV is illustrated in Fig. \ref{fig2}. 

Each subsystem contributes significantly to the overall functionality and safety of the vehicle. The sensors are responsible for collecting real-time environmental data, providing the foundation of perception and navigation. However, failures in this subsystem can affect the vehicle's ability to interpret its surroundings accurately. Sensor failures occur when camera (SF1), LiDAR (SF2), radar (SF3), GPS (SF4), or IMU (SF5) fails due to hardware malfunctions, environmental interference, or signal loss. 

The perception system is responsible for interpreting the environment using sensor data. Failures in this subsystem affect the vehicle's ability to detect and respond to its surroundings. The basic events in the perception system are labeled PF1 through PF14. Data misalignment (PF1), coordinate frame errors (PF2), and algorithm fusion errors (PF3) may cause inconsistencies in sensor fusion. Furthermore, object recognition failures, such as detecting non-existent objects (PF4), failure to detect existing objects (PF5), and misclassification (PF6), result in inaccurate scene interpretation. Low confidence score (PF7) and edge case limitation (PF8) present additional challenges for object recognition algorithms. Object tracking is responsible for continuously monitoring detected objects in the environment and predicting their future position over time. Failures in object tracking, including data association errors (PF9), drift in tracking output (PF10), and tracking loss (PF11), degrade the reliability of AV perception, leading to unsafe driving decisions. Map matching errors (PF12), coordinate transformation failures (PF13), and localization drift (PF14) in localization introduce navigation inconsistencies, affecting AV's positional accuracy. 

\begin{figure*}[htbp]
\centerline{\includegraphics[scale=0.23]{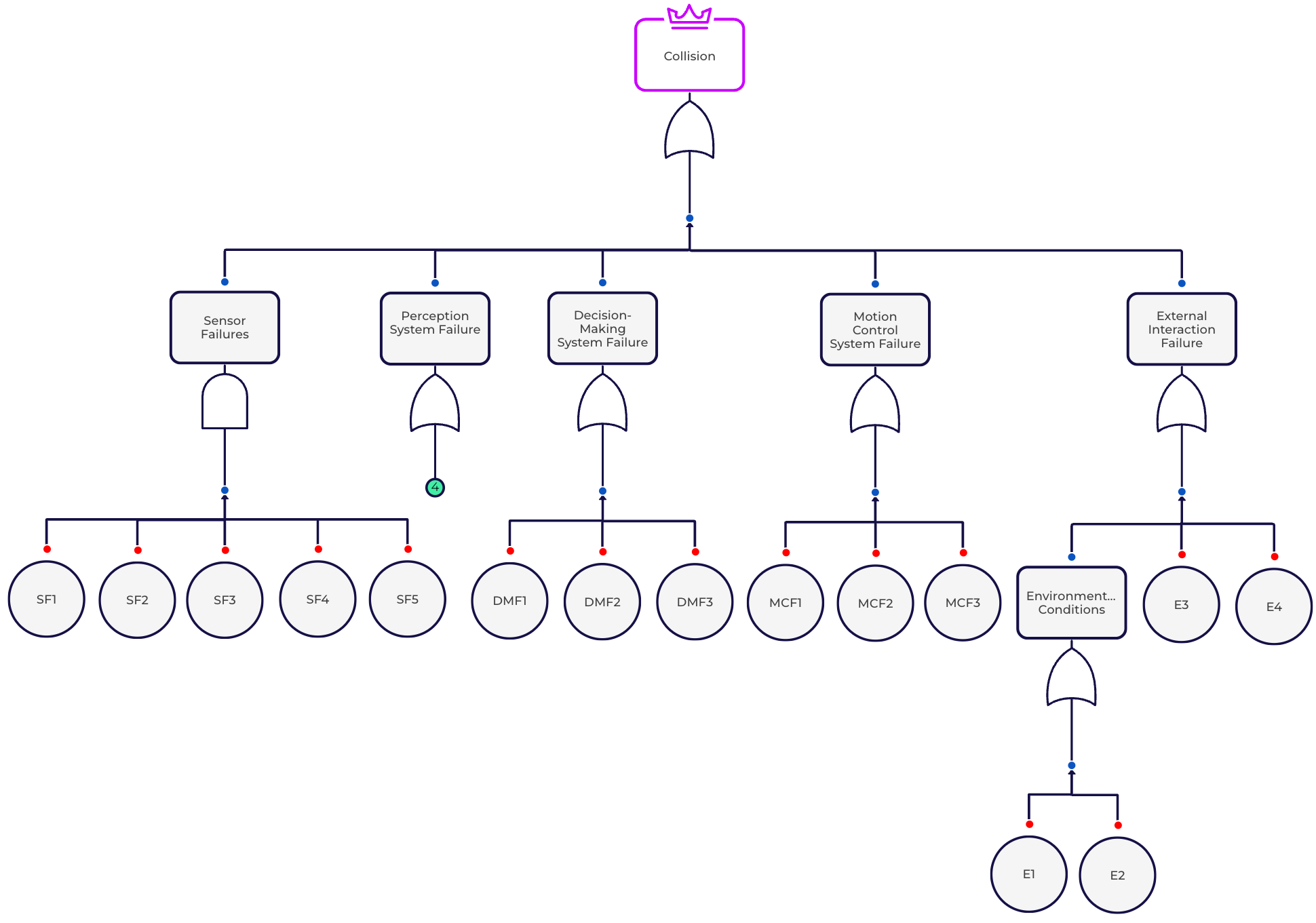}}
\caption{Fault Tree Diagram of Collision as Top Event} 
\label{fig3}
\end{figure*}

\begin{figure*}[htbp]
\centerline{\includegraphics[scale=0.23]{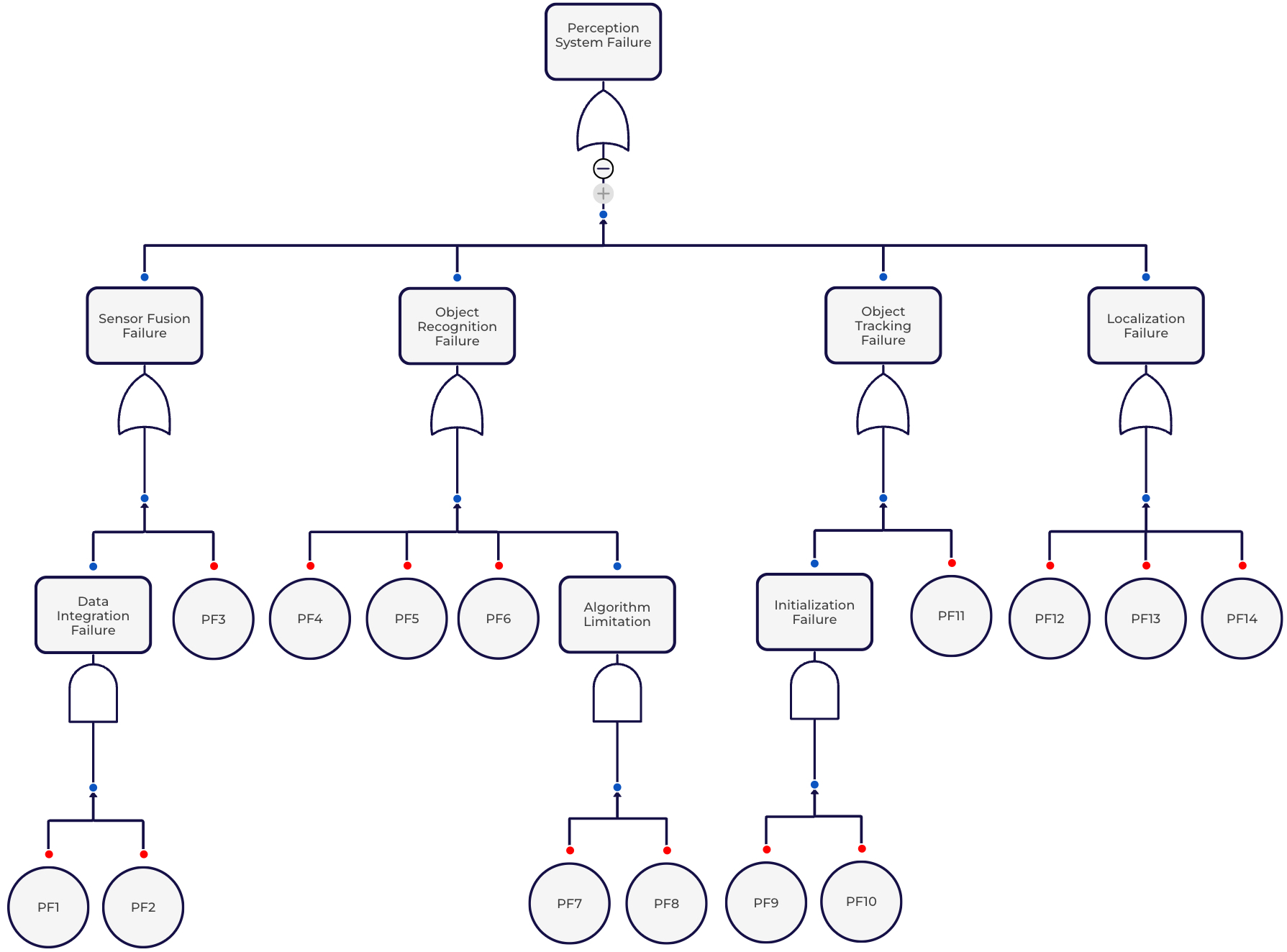}}
\caption{Fault Tree Diagram of Perception System in Collision as Top Event} 
\label{fig4}
\end{figure*}

The decision-making system processes perception outputs and determines appropriate navigation actions. Failures within this subsystem can result in incorrect or delayed driving decisions, increasing the likelihood of collisions. Incorrect path planning (DMF1), delayed response to obstacles (DMF2), and obstacle avoidance failure (DMF3), can directly lead to unsafe driving behavior. The motion control system ensures that planned vehicle actions are executed safely. Failures in acceleration control (MCF1), braking mechanisms (MCF2), or steering functionality (MCF3) affect vehicle stability. In addition to internal system failures, external interaction factors also contribute to AV safety risks. Adverse weather conditions (E1), degraded road conditions (E2), communication failure (E3), and cyberattacks (E4) pose significant challenges. These external factors may reduce sensor accuracy, interfere with decision-making processes, or compromise system security, increasing the likelihood of a collision. Based on the architecture of the autonomous vehicle, we constructed the fault tree with the top event as collision in AVs as Fig. \ref{fig3} and Fig. \ref{fig4}.

\begin{table}
\centering
\renewcommand{\arraystretch}{1.5}
\begin{tabular}{|p{1cm}cp{1.5cm}|}
\hline
\textbf{Event Node} & \textbf{Name of Event} & \textbf{Failure Rate (FIT)} \\
\hline
SF1 & Camera Failure & $6.36 \pm 1.33$  \\\hline
SF2 & LiDAR Failure & $5.67 \pm 1.36$ \\\hline
SF3 & Radar Failure & $6.51 \pm 1.42$ \\\hline
SF4 & GPS Failure & $6.87 \pm 1.74$  \\\hline
SF5 & IMU Failure & $6.46 \pm 1.14 $  \\\hline
PF1 & Data Misalignment & $5.45\pm 0.516$\\\hline
PF2 & Coordinate Frame Errors & $4.82 \pm 0.632$ \\\hline
PF3 & Algorithm Fusion Error & $5.52 \pm 0.806$ \\\hline
PF4 & Detecting Non-Existent Objects & $6.42 \pm 0.843$\\\hline
PF5 & Failure to Detect Existing Objects & $6.46 \pm 0.917$ \\\hline
PF6 & Misclassification & $6.56 \pm 0.454$ \\\hline
PF7 & Low Confidence Scores & $5.01 \pm 0.49$ \\ \hline
PF8 & Edge Case Limitations & $4.27 \pm 1.05$ \\\hline
PF9 & Data Association Errors & $5.43 \pm 0.905$ \\\hline
PF10 & Drift in Tracking Output & $5.17 \pm 0.617$\\ \hline
PF11 & Tracking Loss & $5.01 \pm 0.568$ \\ \hline
PF12 & Map Matching Errors & $5.19 \pm 1.11$ \\ \hline
PF13 & Coordinate Transformation Faults & $5.44 \pm 0.444$ \\ \hline
PF14 & Localization Drift & $5.46 \pm 0.466$ \\ \hline
DMF1 & Incorrect Path Planning & $5.98 \pm 0.787$ \\ \hline
DMF2 & Delayed Response to Obstacle & $6.56 \pm 1.08$ \\ \hline
DMF3 & Obstacle Avoidance Failure & $6.31 \pm 0.923$ \\ \hline
MCF1 & Accelerator Control System Failure & $5.58 \pm 0.775$ \\ \hline
MCF2 & Brake Control System Failure & $5.62 \pm 0.750$ \\ \hline
MCF3 & Steering System Failure & $4.96 \pm 0.812$ \\ \hline
E1 & Weather & $4.48\pm 0.713$ \\ \hline
E2 & Road Conditions & $4.41 \pm 0.607$ \\ \hline
E3 & Communication Failure & $4.76 \pm 0.719$ \\ \hline
E4 & Cyberattack & $5.28 \pm 0.626$ \\
\hline
\end{tabular}
\vspace{0.15cm}
\caption{List of basic events with their corresponding posterior failure rate}
\label{tab:basic_events}
\end{table}

\subsection{Risk-Based Safety Assessment Methodology}
This paper presents a risk-based safety assessment methodology for collision risk analysis in AVs, focusing on defining target failure rates for critical components to align with functional safety standards, particularly ASIL B under ISO 26262. While the ASIL B failure rate target is originally defined for random hardware failures, we use it notionally as a reference for the overall system, including both hardware and software failures, to ensure a comprehensive risk assessment. The methodology systematically analyzes AV architecture by breaking it down into onboard sensors, perception, decision-making, and motion control systems to identify failure components and establish target failure rates. The root causes of collision as the top event are investigated and FT is developed based on literature reviews and expert opinions. In the qualitative analysis, minimal cut sets are generated using FTA to identify the critical failure combinations systematically. For the quantitative analysis, the FT is converted into a BN to perform the probabilistic risk assessment.

\section{Experimental Results}
\subsection{Qualitative Analysis}
The qualitative analysis of the fault tree structure provides insights into the system's minimal cut sets, which represent the simplest combinations of basic events that can cause system failure: collision. Analyzing these cut sets helps to identify which components are the most critical contributors to the risk associated with the system. 

In this study, three order-2 minimal cut sets are identified: coordinate frame errors and data misalignment (PF1, PF2), low confidence scores and edge case limitation (PF7, PF8), and data association errors and drift in tracking outputs (PF9, PF10). These failure pairs indicate that a single failure alone is not sufficient to cause a system failure in these cases. However, the simultaneous occurrence of both failures in a set significantly increases collision risk. To mitigate the risks associated with these minimal cut sets, it needs targeted risk reduction strategies. 

For PF1 and PF2, real-time sensor calibration should be implemented to continuously adjust sensor alignment and correct errors. Besides, sensor fusion consistency checks can detect and mitigate misaligned data before it propagates. For PF7 and PF8, improvements in AI perception systems are necessary. Expanding training datasets with edge-case scenarios can enhance robustness. Adaptive AI models should be developed to handle low-confidence situations by triggering a fail-safe mode or a secondary decision validation process. PF9 and PF10 significantly impact the initialization of object tracking results. Incorporating appearance descriptors can help improve identification matching and reduce correspondence errors. Using probabilistic filters such as Kalman filters, can help reduce drift by estimating covariance and dynamically adjusting prediction confidence to reduce positional uncertainty. In addition, an AND gate connects multiple sensor failures (SF1-SF5). This structure indicates that individual sensor failures do not significantly influence collision risk unless multiple sensors fail simultaneously. This insight suggests that while individual sensor failures are less critical, maintaining overall sensor reliability remains essential to system robustness. 

In addition to the minimal cut sets mentioned above, all other failures in the system are single points of failure, meaning that the failure of a single node directly impacts the probability of collision. These failure modes are particularly high-risk and require additional safety mechanisms to ensure system robustness. Mitigation strategies such as system redundancy (e.g. secondary braking mechanisms, alternative localization) to maintain backup functionality during failures, and AI-driven predictive maintenance to detect early component degradation through real-time operational data analysis. These mechanisms maintain operational safety and reduce catastrophic risk during critical failure.

\subsection{Quantitative Analysis}

\begin{figure*}[htbp]
\centerline{\includegraphics[scale=0.17]{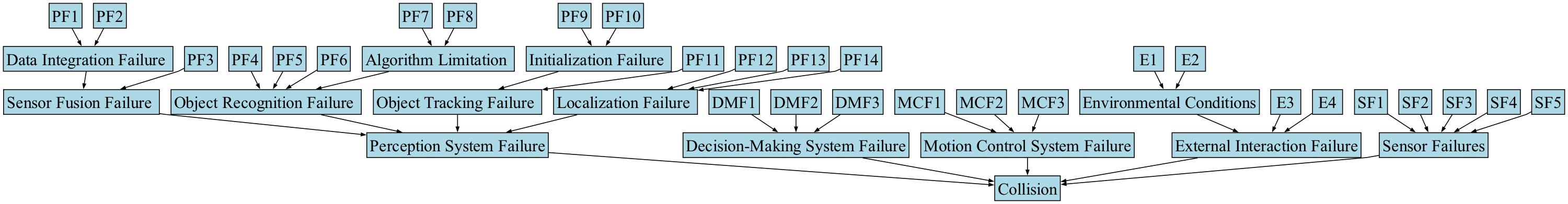}}
\caption{Integration of FT with BN}
\label{fig5}
\end{figure*}

To ensure compliance with ASIL B, we assume that the failure rate of a collision in AV systems does not exceed 100 FIT. FIT stands for failures in time which is a unit used to express the failure rate of a component or system. Following the fault tree structure, we randomly assign failure rates to basic events while ensuring the total failure rate remains on this predefined safety threshold. The fault tree is constructed and evaluated using Pathfinder, and failure probabilities for basic events are calculated using the formula: \(P=1-e^{- \lambda t}\) where \(P\) denotes the failure probability, \(\lambda\) represents the failure rate, and \(t\) stands for the time. In our study, \(t\) is set to 10,000 operational hours since we are working on an autonomous system. Then these probabilities are mapped into a BN for posterior probability estimation.

\begin{figure}[htbp]
\centerline{\includegraphics[scale=0.4]{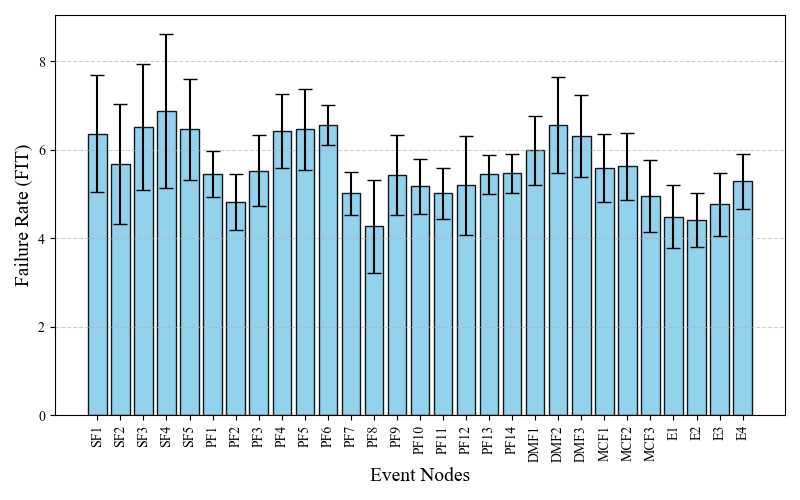}}
\caption{Failure Rates of Basic Events}
\label{fig6}
\end{figure}

Following the conversion rules from FTA to BN, we construct a Bayesian Network model using Python, as shown in Fig. \ref{fig5}. To validate the BN structure, we assume a collision failure probability of 0.001 and ensure the BN reproduces the same probability. The probability of collision calculated in both FTA and BN is identical at 0.001, confirming the consistency and accuracy of the two models. Table \ref{tab:basic_events} and Fig. \ref{fig6} present the failure rates of the basic events with 95\% confidence levels. These provide a refined understanding of the system's vulnerabilities and highlight critical components contributing to the overall collision risk. 

Among the analyzed basic events, several basic events are identified as high-risk contributors. PF5 (\(6.42 \pm 0.843\) FIT) and PF6 (\(6.56\pm 0.454\) FIT) highlight object detection algorithm limitations in perception system, particularly under adverse conditions. The failure rate of DMF2 is \(6.56 \pm 1.08\) FIT, indicating that latencies in decision-making processes substantially increase collision risk. Similarly, DMF3 demonstrated a failure rate of \(6.31 \pm 0.923\) FIT, highlighting the need for enhanced decision-making systems to handle complex driving scenarios effectively. Beyond internal system failures, E4 emerges as a significant external failure event, highlighting the growing risk of cybersecurity threats to autonomous vehicles. Cyberattacks have the potential to interfere with perception and decision-making algorithms, compromise vehicle control systems, and cause incorrect actions, which raises the collision risk.

Some events show high variability in failure rates, such as PF8 (\(4.27 \pm 1.05\) FIT), indicating significant uncertainty in handling rare or extreme scenarios. This highlights the need for better AI training on diverse datasets to address edge-case performance issues. Moreover, the sensor subsystem shows particularly high variability in failure rates, which can be attributed to factors such as environmental influences, calibration issues, or gradual degradation of sensor components over time. Since the AND gate is used to model sensor dependencies, the overall probability of sensor subsystem failure remains relatively low. To mitigate these uncertainties and improve system robustness, it is essential to implement redundant sensor configurations, advanced sensor fusion methodologies, and real-time diagnostic mechanisms. These measures will enhance fault tolerance and ensure more reliable perception performance under diverse operating conditions.

The perception system (40.06 FIT) is the most significant contributor to collision risk, followed by decision-making (18.85 FIT), motion control (16.16 FIT), and external interaction subsystems (18.93 FIT). By enhancing object detection accuracy and expanding the training datasets with more diverse and representative scenarios, the system can better handle edge cases and adverse conditions, such as extreme weather or poor visibility. These enhancements will help mitigate failures such as PF5, and PF6, which are critical contributors to the overall collision risk. A well-trained perception system will reduce the likelihood of misidentification and failure to detect objects, especially in challenging environments. 

The implementation of real-time validation mechanisms for cross-checking data from multiple sensors can greatly reduce sensor fusion failures. This approach is particularly effective in preventing failures PF1 by ensuring that inconsistencies in data alignment are detected and corrected. Moreover, introducing redundancy in perception systems helps mitigate single-sensor malfunctions, preventing cascading failures that could disrupt object tracking and motion planning. This approach is particularly valuable for mitigating failures such as PF9 and PF11, which can disrupt the tracking and continuity. In the decision-making subsystem, failures such as DMF3 can be mitigated by using predictive decision algorithms and fail-safe override mechanisms. These enhancements allow the system to anticipate dynamic obstacles and respond immediately to avoid collisions. For the motion control subsystem, which includes components like brakes and steering, redundancy is critical. Introducing backup control systems and deploying predictive maintenance can reduce risks associated with MCF2 and MCF3. The external interaction subsystem, which includes environmental factors like E2, can benefit from adaptive driving algorithms that dynamically adjust to varying road conditions, minimizing their impact on vehicle safety. By focusing on specific failure events using these mitigation techniques, the subsystems can operate more effectively, improving the safety and robustness of autonomous vehicles.

\section{Conclusion}
In this paper, we have presented a risk-based safety assessment methodology for AV collision risk analysis by integrating FTA and BN. Traditional FTA provides a structured breakdown of failure events but assumes static failure probabilities and lacks the ability to model interdependencies. To address this limitation, we have applied Bayesian inference to refine failure rates and establish ASIL B target failure rates for critical AV components. We construct a fault tree based on AV subsystem architecture, defining collision as the top event. Failure rates are assigned while ensuring the collision remained within 100 FIT, which is the predefined safety threshold. By converting to BN, Bayesian inference is used to update posterior probabilities, repeated over 10 times and analyzed at a 95\% confidence level. Results indicate that perception system failures (46.06 FIT) contribute most to collision risk, followed by decision-making (18.85 FIT), motion control (16.16 FIT), and external interactions (18.93 FIT). Delayed response to obstacles (DMF2), misclassification (PF6), and failure to detect existing objects (PF5) are identified as critical failure points. Mitigation strategies are proposed for each subsystem, focusing on improving sensor redundancy, enhancing perception algorithms, and optimizing decision-making processes. The integration of FTA and BN enables dynamic risk quantification, providing system designers with refined failure rate targets. Future work will explore real-time Bayesian inference implementation and enhanced environmental modeling to improve AV safety assessment.


\section*{Acknowledgment}
This work is supported in part by MITACS Accelerate Program project IT40981, NSERC CREATE TRAVERSAL program, Ontario Research Fund-Research Excellence (ORF-RE) program under RE012-026, and in part by reasonX Labs. We would like to acknowledge reasonX Labs team for valuable discussions and support during this work.

\bibliographystyle{IEEEtran}

\end{document}